\documentclass[12pt,a4paper]{article}
\usepackage{amsmath}
\usepackage{amssymb}
\usepackage[cp866]{inputenc}
\usepackage[russian]{babel}

\begin{document}

\begin{center}
\textbf{Study of the Low-Energy Characteristics of Neutron-Neutron
Scattering in the Effective-Range Approximation}\bigskip

\textbf{V. A. Babenko\footnote{%
E-mail: pet2@ukr.net} and N. M. Petrov}

\bigskip

\textit{Bogolyubov Institute for Theoretical Physics, National Academy of
Sciences of Ukraine, Kiev}
\end{center}

\thispagestyle{empty}\bigskip

\vspace{1pt}

\noindent The influence of the mass difference between the charged and
neutral pions on the low-energy characteristics of nucleon-nucleon
interaction in the $^{1}S_{0}$ spin-singlet state is studied within the
framework of the effective-range approximation. By making use of the
experimental singlet neutron-proton scattering parameters and the
experimental value of neutron-neutron virtual-state energy, the following
values were obtained for the neutron-neutron scattering length and effective
range: $a_{nn}=-16.59(117)\,$fm, $r_{nn}=2.83(11)\,$fm. The calculated
neutron-neutron scattering length $a_{nn}$ is in good agreement with one of
the two well known and differing experimental values of this quantity, and
the calculated effective range $r_{nn}$ is also in good agreement with
present-day experimental results.\bigskip \bigskip

\bigskip \bigskip

\begin{center}
1. INTRODUCTION\bigskip
\end{center}

Low-energy parameters of nucleon-nucleon ($NN$) interaction are fundamental
quantities that play a significant role in studying strong nucleon-nucleon
interaction [1--15]. These quantities are also of importance in constructing
various nuclear-force models, which, in turn, form a basis for exploring the
structure and properties of nuclei, as well as various nuclear processes
[3--10, 16--20].

Investigations of low-energy parameters of nucleon-nucleon interaction in
the spin-singlet $^{1}S_{0}$ state are of particular importance in
connection with the problem of testing the hypothesis of charge independence
and charge symmetry of nuclear forces [4, 6, 10, 16--18]. This hypothesis is
confirmed to some extent by an approximate equality of binding energies of
isobar nuclei. However, experimental and theoretical investigations aimed at
quite a precise determination of low-energy parameters for neutron-proton ($%
np$), proton-proton ($pp$), and neutron-neutron ($nn$) interactions are of
paramount importance for definitively solving this problem.

The parameters of neutron-proton and proton-proton systems can be determined
to a high degree of precision directly from experiments. At the same time,
it is impossible to study neutron-neutron scattering directly because of the
absence of neutron targets. In order to determine experimentally the
parameters of neutron-neutron interaction, use is usually made of nuclear
reactions leading to the production of two interacting neutrons in the final
state. A detailed review of experimental and theoretical methods for
determining the neutron-neutron scattering length $a_{nn}$ from data on such
reactions was given in [18]. On the basis of a vast body of experimental
data obtained before 1974, a mean-weighted value of $a_{nn}=-16.61(54)\,$fm
was found in [18] for the neutron-neutron scattering length. It is
noteworthy that this value is consistent with present-day data.

The deuteron-breakup reaction $n+d\longrightarrow p+n+n$ is one of the
reactions leading to the appearance of two interacting neutrons in the final
state. A fit to data on this reaction on the basis of the Migdal-Watson
formula [21, 22] makes it possible to determine the value of the
virtual-state energy $\varepsilon _{nn}$ for the neutron-neutron system in
the $^{1}S_{0}$ state [23--25].

At the present time, a description of nucleon-nucleon interaction frequently
relies on semiphenomenological one-boson-exchange potential models involving
the exchange of various mesons. Within this approach, the exchange of pions,
which are light, determines primarily the long-range part of the
nucleon-nucleon potential, while the exchange of heavier rho and omega
mesons determines the interaction at intermediate and short distances, which
is significant at higher energies. At extremely low energies, which
effectively correspond to long distances, the use of extremely simple
one-pion-exchange potentials in describing nucleon-nucleon interaction is
quite reasonable. Taking into account the foregoing and assuming that
nuclear forces in the nucleon-nucleon system at low energies are due
primarily to the exchange of virtual pions, we study here, in the
effective-range approximation, the influence of the mass difference between
the charged and neutral pions on low-energy parameters of nucleon-nucleon
scattering in the $^{1}S_{0}$ state. We find a relation between low-energy
parameters of the neutron-proton system and their counterparts for the
neutron-neutron system. The use of this relation and the energy of a virtual
level in the neutron-neutron system makes it possible to determine the
scattering length $a_{nn}$ and the effective range $r_{nn}$.\bigskip

\bigskip

\begin{center}
2. BASIC RELATIONS OF THE EFFECTIVE-RANGE APPROXIMATION\bigskip
\end{center}

It is important to investigate the low-energy parameters of nucleon-nucleon
interaction in the spin-singlet $^{1}S_{0}$ state in view of the need for
testing the hypothesis of charge independence and charge symmetry of nuclear
forces. The effective-range approximation 
\begin{equation}
k\cot \delta _{NN}=-\frac{1}{a_{NN}}+\frac{1}{2}r_{NN}k^{2},  \tag{1}
\end{equation}%
where $a_{NN}$ and $r_{NN}$ are, respectively, the scattering length and the
effective range for nucleon-nucleon interaction, is the most convenient and
the most popular method for analyzing experimental data on nucleon-nucleon
scattering at low energies [1--15]. If use is made of the approximation
specified by Eq. (1), the $S$ matrix can be written in the form%
\begin{equation}
S\left( k\right) =\frac{k+i\alpha _{NN}}{k-i\alpha _{NN}}~\frac{k+i\beta
_{NN}}{k-i\beta _{NN}}~,  \tag{2}
\end{equation}%
where%
\begin{equation}
\alpha _{NN}=\frac{1}{r_{NN}}\left[ 1-\left( 1-\frac{2r_{NN}}{a_{NN}}\right)
^{1/2}\right] ,  \tag{3}
\end{equation}%
\begin{equation}
\beta _{NN}=\frac{1}{r_{NN}}\left[ 1+\left( 1-\frac{2r_{NN}}{a_{NN}}\right)
^{1/2}\right] .  \tag{4}
\end{equation}

The $S$ matrix in (2) has two poles in the complex plane of the wave number $%
k$. The pole $i\alpha _{NN}$, situated in the lower half-plane of $k$,
corresponds to a virtual ($\alpha _{NN}<0$) state of the two-nucleon system
at the energy%
\begin{equation}
\varepsilon _{NN}=\frac{\hbar ^{2}\alpha _{NN}^{2}}{2m_{NN}},  \tag{5}
\end{equation}%
where $m_{NN}$ is the reduced mass of the two-nucleon system. The second
pole $i\beta _{NN}$ ($\beta _{NN}>0$), situated in the upper half-plane of $%
k $, is the well-known redundant pole of the $S$ matrix.

The potential corresponding to the effective-range approximation (1) has the
form [5, 26, 27]%
\begin{equation}
V_{NN}\left( r\right) =-V_{NN0}\frac{e^{-r/R_{NN}}}{\left( 1+\gamma
_{NN}e^{-r/R_{NN}}\right) ^{2}},  \tag{6}
\end{equation}%
where%
\begin{equation}
R_{NN}=\frac{1}{2\beta _{NN}},  \tag{7}
\end{equation}%
\begin{equation}
V_{NN0}=\frac{\hbar ^{2}}{m_{NN}}\frac{\gamma _{NN}}{R_{NN}^{2}},  \tag{8}
\end{equation}%
\begin{equation}
\gamma _{NN}=\frac{1+2\alpha _{NN}R_{NN}}{1-2\alpha _{NN}R_{NN}}.  \tag{9}
\end{equation}%
At large distances, the potential in (6) decreases exponentially; that is,%
\begin{equation}
V_{NN}\left( r\right) \underset{r\rightarrow \infty }{\simeq }%
-V_{NN0}e^{-r/R_{NN}}.  \tag{10}
\end{equation}%
As can be seen from Eq. (7), the potential range $R_{NN}$ is determined
directly in terms of the parameter $\beta _{NN}$, which characterizes the
redundant pole of the $S$ matrix in (2). In the case of the approximation
specified by Eq. (1), Eqs. (3) and (4) can be used together with Eq. (7) to
derive equations relating low-energy parameters of the nucleon-nucleon
system to the interaction range $R_{NN}$ as%
\begin{equation}
r_{NN}=4R_{NN}\left( 1-\frac{2R_{NN}}{a_{NN}}\right) ,  \tag{11}
\end{equation}%
\begin{equation}
r_{NN}=4R_{NN}/\left( 1+2\alpha _{NN}R_{NN}\right) .  \tag{12}
\end{equation}%
Relations (11) and (12), together with the values of the interaction range $%
R_{NN}$ and virtual-state energy (5), make it possible to determine the
nucleon-nucleon scattering length $a_{NN}$ and effective range $r_{NN}$%
.\bigskip \bigskip

\begin{center}
3. DETERMINATION OF LOW-ENERGY PARAMETERS OF PROTON-PROTON AND
NEUTRON-NEUTRON SCATTERING\bigskip
\end{center}

A comparison of the scattering lengths and effective ranges for
neutron-proton, proton-proton, and neutron-neutron scattering is one of the
methods that makes it possible to verify the hypotheses of charge
independence and charge symmetry of nuclear forces. In this case, one can
also compare the values of the virtual-state energies in the aforementioned
nucleon-nucleon systems. In comparing low-energy parameters of
nucleon-nucleon interaction, one should bear in mind that, in the case of
proton-proton interaction, it is necessary to use purely nuclear low-energy
parameters obtained by excluding the electromagnetic component of the
proton-proton interaction from the experimental values of the parameters in
question.

According to Yukawa's meson theory, strong nuclear interaction between two
nucleons is due largely to the exchange of virtual pions, which determines
the long-range part of the nucleon-nucleon interaction and, accordingly,
scattering at extremely low energies. The respective nuclear-force range $R$
is in inverse proportion to the pion mass and is small; that is, 
\begin{equation}
R\simeq \frac{\hbar }{m_{\pi }c}\approx 1.4~\text{fm}.  \tag{13}
\end{equation}%
The reason why the nuclear-force range takes different values in the case of
two neutrons (or two protons) and in the neutron-proton case is the
following: identical nucleons exchange a neutral pion of mass%
\begin{equation}
m_{\pi ^{0}}=134.9766(6)~\text{MeV},  \tag{14}
\end{equation}%
while a neutron and a proton may exchange both a neutral pion ($\pi ^{0}$)
and a charged pion ($\pi ^{\pm }$) of mass%
\begin{equation}
m_{\pi ^{\pm }}=139.57018(35)~\text{MeV}.  \tag{15}
\end{equation}%
In view of this, it is natural to assume that%
\begin{equation}
R_{nn}=R_{pp}=\frac{\overline{m}_{\pi }}{m_{\pi ^{0}}}R_{np},  \tag{16}
\end{equation}%
where $\overline{m}_{\pi }\equiv \left( 2m_{\pi ^{\pm }}+m_{\pi ^{0}}\right)
/3$ is the average mass of the neutral and charged pions, the ratio of the
average pion mass $\overline{m}_{\pi }$ to the neutral-pion mass being%
\begin{equation}
\frac{\overline{m}_{\pi }}{m_{\pi ^{0}}}=1.0227.  \tag{17}
\end{equation}

In determining and comparing the scattering lengths and effective ranges for
the nucleon-nucleon systems in the $^{1}S_{0}$ spin-singlet state, we will
use the singlet neutron-proton scattering length $a_{np}$ and effective
range $r_{np}$ as reference values and take their experimental values from
[4, 28], where they were%
\begin{equation}
a_{np}=-23.71(1)\,\text{fm},  \tag{18}
\end{equation}%
\begin{equation}
r_{np}=2.70(9)\,\text{fm}.  \tag{19}
\end{equation}%
With allowance for the experimental value of the deuteron binding energy
[29],%
\begin{equation}
\varepsilon _{d}=2.224575(9)\,\text{MeV},  \tag{20}
\end{equation}%
and the experimental value of the neutron-proton coherent scattering length
[30, 31],%
\begin{equation}
f=-3.7406(11)\,\text{fm},  \tag{21}
\end{equation}%
the parameter values in (18) and (19) make it possible to describe, to a
high precision, experimental data on the scattering of neutrons by protons
in the region of low energies. We note that the values in (18) and (19) used
for the singlet low-energy parameters of neutron-proton scattering are in
excellent agreement with their values of $a_{np}=-23.7154(80)\,$fm and $%
r_{np}=2.706(67)$~fm recently obtained by the present authors in [14] on the
basis of modern experimental data.

Starting from the parameter values in (18) and (19) and using Eqs. (4) and
(7), we find for the nuclear-forces range in the neutron-proton system that%
\begin{equation}
R_{np}=0.6404(204)\,\text{fm}.  \tag{22}
\end{equation}%
With allowance for this value and according to Eqs. (16) and (17), we obtain
the following results for the nuclear-force ranges in the neutron-neutron
and proton-proton systems:%
\begin{equation}
R_{nn}=R_{pp}=0.6549(208)\,\text{fm}.  \tag{23}
\end{equation}

By using the value of the purely nuclear proton-proton scattering length
from [4],%
\begin{equation}
a_{pp}=-17.2(1)\,\text{fm},  \tag{24}
\end{equation}%
in considering the effective range for proton-proton scattering, we deduce
from Eqs. (11) and (23) the following value for the purely nuclear effective
range in proton-proton scattering:%
\begin{equation}
r_{pp}=2.819(97)\,\text{fm}.  \tag{25}
\end{equation}%
This value agrees well with the currently recommended value [6]%
\begin{equation}
r_{pp}=2.85(4)\,\text{fm}.  \tag{26}
\end{equation}

In order to determine the low-energy parameters of neutron-neutron
scattering, we employ the experimental results obtained recently in [23--25]
for the $nd$ breakup reaction. In [23--25], the experimental dependence of
the yield of the reaction $n+d\longrightarrow p+n+n$ on the energy of
relative motion of two final-state neutrons, $\varepsilon $, was simulated
on the basis of the Migdal-Watson formula [21, 22]%
\begin{equation}
F_{MW}\left( \varepsilon \right) =A\frac{\sqrt{\varepsilon }}{\varepsilon
+\varepsilon _{nn}},  \tag{27}
\end{equation}%
where the parameter $A$ and the energy of the virtual state in the
neutron-neutron system, $\varepsilon _{nn}$, are adjustable parameters. A
simulation of the $nd$ breakup reaction by using Eq. (27) makes it possible
to determine the virtual-state energy $\varepsilon _{nn}$ directly from
experimental data. This energy is related to the scattering length $a_{nn}$
and the effective range $r_{nn}$ by the equation%
\begin{equation}
\frac{1}{a_{nn}}=-\left( \frac{m_{n}\varepsilon _{nn}}{\hbar ^{2}}\right)
^{1/2}-\frac{1}{2}r_{nn}\frac{m_{n}\varepsilon _{nn}}{\hbar ^{2}},  \tag{28}
\end{equation}%
where $m_{n}$ is the neutron mass. In terms of the virtual-state wave number 
$\alpha _{nn}$, this equation assumes the form%
\begin{equation}
\frac{1}{a_{nn}}=\alpha _{nn}-\frac{1}{2}r_{nn}\alpha _{nn}^{2}.  \tag{29}
\end{equation}

The value deduced in [23--25] for the neutron-neutron scattering length from
a comparison of the experimental data and the results obtained by simulating
the reaction $n+d\longrightarrow p+n+n$ on the basis of the Migdal-Watson
formula (27) in the zero-range approximation for nuclear forces ($r_{nn}=0$)
is%
\begin{equation}
a_{nn}=-17.9(10)\,\text{fm}.  \tag{30}
\end{equation}%
The following values of the virtual-state energy in the neutron-neutron
system and of the wave number correspond to the $a_{nn}$ value in (30):%
\begin{equation}
\varepsilon _{nn}=0.1293(158)\,\text{MeV},  \tag{31}
\end{equation}%
and%
\begin{equation}
\alpha _{nn}=-0.0559(33)\,\text{fm}^{-1}.  \tag{32}
\end{equation}%
Below, we will use the values in (31) and (32) as experimental values of the
energy and wave number of the virtual state in the neutron-neutron system.

Further, we take into account the finiteness of the nuclear-force range and
consider neutron-neutron interaction in the effective-range approximation
(1), which is more precise than the zero-range approximation. By using Eqs.
(12), (23), and (32), we obtain the following value for the effective range
in neutron-neutron scattering: 
\begin{equation}
r_{nn}=2.827(111)\,\text{fm}.  \tag{33}
\end{equation}%
From Eqs. (29), (32), and (33), we find that the neutron-neutron scattering
length $a_{nn}$ is%
\begin{equation}
a_{nn}=-16.59(117)\,\text{fm}.  \tag{34}
\end{equation}%
Thus, we see that, upon taking into account the finiteness of the
nuclear-force range, the absolute value of the neutron-neutron scattering
length decreases in relation to what we have in the zero-range approximation
by the value%
\begin{equation}
\Delta a_{nn}\simeq 1.3\,\text{fm},  \tag{35}
\end{equation}%
which is $7.3\%$ in relative units. We also note that, with the aid of Eqs.
(23) and (32), the value identical to that in (34) can be obtained for the
neutron-neutron scattering length from the equation%
\begin{equation}
a_{nn}=\frac{1}{\alpha _{nn}}+2R_{nn}~.  \tag{36}
\end{equation}%
Equation (36) follows directly from Eqs (11) and (12).

The table gives values of basic low-energy parameters of nucleon-nucleon
interaction in the $^{1}S_{0}$ state. One can see from the table that the
values of the low-energy parameters of neutron-proton interaction differ
substantially from their counterparts in the cases of proton-proton and
neutron-neutron interaction. The difference in the scattering length is
about $7$ fm, which is $30\%$ in relative units. The values of the
virtual-state energy in the proton-proton and neutron-neutron systems exceed
the virtual-state energy in the neutron-proton system by $80\%$ and $95\%$,
respectively.

The effective ranges in proton-proton and neutron-neutron scattering are
approximately equal to each other and exceed the effective range in
neutron-proton scattering by $4.8\%$. This is due to the mass difference
between the neutral and charged pions. It is noteworthy that the
nuclear-force range $R_{NN}$ governs $90\%$ of the effective ranges for
nucleon-nucleon scattering. Thus, the approximate equality of the effective
ranges for proton-proton and neutron-neutron scattering is due to the
equality of the nuclear-force ranges in proton-proton and neutron-neutron
interactions ($R_{nn}=R_{pp}=0.6549\,$fm). Since the nuclear-force range in
proton-proton and neutron-neutron interactions is longer than the
nuclear-force range in neutron-proton interaction ($R_{np}=0.6404\,$fm), the
effective ranges for proton-proton and neutron-neutron scattering exceed the
effective range for neutron-proton scattering by about $0.12$ fm.

From Eq. (11), it follows that the effective ranges for proton-proton ($%
r_{pp}$) and neutron-neutron ($r_{nn}$) scattering are weakly sensitive to
variations in the scattering lengths $a_{pp}$ and $a_{nn}$; that is,%
\begin{equation}
\Delta r_{NN}=8\frac{R_{NN}^{2}}{a_{NN}^{2}}\Delta a_{NN},  \tag{37}
\end{equation}%
where $8R_{NN}^{2}/a_{NN}^{2}\approx 0.012$ for proton-proton and
neutron-neutron scattering and $8R_{NN}^{2}/a_{NN}^{2}\approx 0.006$ for
neutron-proton scattering. In particular, the effective range $r_{nn}$
changes from $2.791$ to $2.848$ fm in response to the change in the
neutron-neutron scattering length $a_{nn}$ from $-20$ to $-15$ fm. The
effective ranges for proton-proton and neutron-neutron scattering are also
weakly sensitive to variations in the virtual-state wave number $\alpha
_{NN} $ and, hence, to variations in the virtual-state energy itself.
Therefore, the error in the effective ranges for proton-proton and
neutron-neutron scattering is determined primarily by the error in the
nuclear-force range and is virtually coincident with the error in the
effective range for neutron-proton scattering.

Thus, it follows from a comparison of the low-energy parameters for the
neutron-proton system and their counterparts for the proton-proton and
neutron-neutron systems that the charge independence of nuclear forces is
violated, which is associated with the mass difference between the charged
and neutral pions. As can be seen from the table, it can hardly be concluded
that the charge symmetry of nuclear forces is also violated since the errors
in the scattering length and virtual-state energy in the neutron-neutron
system are overly large. The improvement of precision in experiments devoted
to determining the virtual-state energy in the neutron-neutron system would
make it possible to draw more reliable conclusion on the occurrence of this
violation.\bigskip \bigskip

\begin{center}
4. CONCLUSION\bigskip
\end{center}

The effect of the mass difference between the charged and neutral pions on
the low-energy parameters of nucleon-nucleon interaction in the $^{1}S_{0}$
state has been studied under the assumption that nuclear forces in the
nucleon-nucleon system at low energies are generated primarily by the
exchange of virtual pions. A relation between the low-energy parameters of
the neutron-proton system and their counterparts for the neutron-neutron
system has been found. Using this relation and the value of the
virtual-state energy in the neutron-neutron system, one can determine the
scattering length $a_{nn}$ and the effective range $r_{nn}$ for this system.

As a basis for calculating and comparing the parameters of the proton-proton
and neutron-neutron systems, we have taken the following low-energy
parameters for neutron-proton scattering: $a_{np}=-23.71(2)\,$fm and $%
r_{np}=2.70(9)\,$fm. Using the value of $a_{pp}=-17.2(1)\,$fm extracted from
experimental data for the purely nuclear proton-proton scattering length, we
have calculated the effective nuclear range for proton-proton scattering.
The resulting value of $r_{pp}=2.82(10)\,$fm is in very good agreement with
the currently recommended value of $r_{pp}=2.85(4)\,$fm [6].

By using experimental results from [23--25] on the reaction $%
n+d\longrightarrow p+n+n$, we have calculated low-energy parameters of
neutron-neutron scattering. The results are $a_{nn}=-16.59(117)\,$fm and $%
r_{nn}=2.83(11)\,$fm. The first value agrees very well with the
mean-weighted value of $a_{nn}=-16.61(54)\,$fm from [18] and with the values
of $a_{nn}=-16.5(9)\,$fm and $a_{nn}=-16.2(4)\,$fm from [32] and [33],
respectively. At the same time, this value does not agree with the values of 
$a_{nn}=-18.7(7)\,$fm and $a_{nn}=-18.50(53)\,$fm from [34] and [35],
respectively.

The effective range calculated for neutron-neutron scattering, $%
r_{nn}=2.83(11)\,$fm, is nearly equal to the purely nuclear effective range
for proton-proton scattering, $r_{pp}=2.82(10)\,$fm. Very close agreement of
the values of the effective ranges for proton-proton and neutron-neutron
scattering is due primarily to the equality of the nuclear-force ranges in
proton-proton and neutron-neutron interaction, while their small difference
of about $0.01\,$fm stems, in accordance with Eq. (37), from the difference
of about $0.6\,$fm between the scattering lengths $a_{pp}$ and $a_{nn}$. The
error in the effective ranges $r_{pp}$ and $r_{nn}$ is dominated by the
error in the effective range for neutron-proton scattering. The value of $%
r_{nn}=2.83(11)\,$fm obtained here for the neutron-neutron effective range
on the basis of Eqs. (12), (23), and (32) is in perfect agreement with the
value of $r_{nn}=2.83(11)\,$fm determined directly from the experiment in
[36] and complies well with the value of $r_{nn}=2.85(60)\,$fm from [37].
However, our result in question is in somewhat poorer agreement with the
value of $r_{nn}=2.69(27)\,$fm from [38] and the value of $r_{nn}=2.9(4)\,$%
fm from [39], but remains compatible with them.\bigskip

\begin{center}
REFERENCES\bigskip
\end{center}

\begin{enumerate}
\item J. M. Blatt and J. D. Jackson, Phys. Rev. \textbf{76}, 18\ (1949).

\item H. A. Bethe, Phys. Rev. \textbf{76}, 38\ (1949).

\item L. Hulth\'{e}n and M. Sugawara, in \textit{Handbuch der Physik, }Vol.
39: \textit{Structure of Atomic Nuclei}, Ed. by S. Fl\"{u}gge (Springer, New
York, Berlin, 1957).

\item A. G. Sitenko and V. K. Tartakovskii, \textit{Lectures on the Theory
of the Nucleus }(Pergamon, Oxford, 1975).

\item R. G. Newton, \textit{Scattering Theory of Waves and Particles}
(Springer, New York, Berlin, 1982).

\item G. A. Miller, B. M. K. Nefkens, and I. \v{S}laus, Phys. Rep. \textbf{%
194}, 1\ (1990).

\item V. V. Pupyshev,\textit{\ }Phys. Part. Nucl. \textbf{28}, 1457 (1997).

\item V. A. Babenko and N. M. Petrov, Phys. Atom. Nucl. \textbf{63}, 1709\
(2000).

\item V. A. Babenko and N. M. Petrov, Phys. Atom. Nucl. \textbf{64}, 233\
(2001); V. A. Babenko and N. M. Petrov, arXiv:nucl-th/0307090 (2003).

\item R. Machleidt and I. Slaus,\textit{\ }J. Phys. G. \textbf{27}, R69
(2001).

\item R. W. Hackenburg,\textit{\ }Phys. Rev. C \textbf{73}, 044002 (2006).

\item V. A. Babenko and N. M. Petrov, Phys. Atom. Nucl. \textbf{69}, 1552\
(2006).

\item V. A. Babenko and N. M. Petrov, Phys. Atom. Nucl. \textbf{70}, 669\
(2007); V. A. Babenko and N. M. Petrov, arXiv:0704.1024 [nucl-th] (2007).

\item V. A. Babenko and N. M. Petrov, Phys. Atom. Nucl. \textbf{73}, 1499\
(2010).

\item R. N. Perez, J. E. Amaro, and E. R. Arriola, arXiv:1410.8097 [nucl-th]
(2014).

\item A. Bohr and B. R. Mottelson, \textit{Nuclear Structure, }Vol. 1\textit{%
\ }(Benjamin, New York, 1969).

\item H. Frauenfelder and E. M. Henley, \textit{Subatomic Physics }%
(Prentice-Hall, Englewood Cliffs, New Jersey, 1974).

\item B. K\"{u}hn, Fiz. Elem. Chast. Atom. Yadra \textbf{6}, 347 (1975);
Sov. J. Part. Nucl. \textbf{6}, 139 (1975).

\item V. A. Babenko and N. M. Petrov, Phys. Atom. Nucl. \textbf{71}, 1730\
(2008).

\item M. Naghdi,\textit{\ }Phys. Part. Nucl. \textbf{45}, 924 (2014).

\item A. B. Migdal, Sov. Phys. JETP \textbf{1}, 2 (1955).

\item K. M. Watson, Phys. Rev. \textbf{88}, 1163\ (1952).

\item S. V. Zuev, E. S. Konobeevskii, M. V. Mordovskoi, and S. I. Potashev,
Bull. Russ. Acad. Sci. Phys. \textbf{73}, 796 (2009).

\item E. S. Konobeevskii, Yu. M. Burmistrov, S. V. Zuev, \textit{et al.},
Phys. Atom. Nucl. \textbf{73}, 1302\ (2010).

\item E. S. Konobeevskii, V. I. Kukulin, M. V. Mordovskoi, \textit{et al.},
Bull. Russ. Acad. Sci. Phys. \textbf{75}, 443 (2011).

\item R. G. Newton and T. Fulton, Phys. Rev. \textbf{107}, 1103\ (1957).

\item N. M. Petrov and E. V. Tartakovskaya, Izv. Akad. Nauk SSSR, Ser. fiz. 
\textbf{48}, 1978 (1984).

\item W. O. Lock and D. F. Measday, \textit{Intermediate Energy Nuclear
Physics} (Methuen, London, 1970).

\item C. Van Der Leun and C. Anderliesten, Nucl. Phys. A \textbf{380, }261
(1982).

\item L. Koester and W. Nistler, Z. Phys. A\textbf{\ 272}, 189 (1975).

\item V. F. Sears, Z. Phys. A\textbf{\ 321}, 443 (1985).

\item W. von Witsch, X. Ruan, and H. Wita\l a, Phys. Rev. C \textbf{74},
014001\ (2006).

\item V. Huhn, L. W\"{a}tzold, C. Weber, \textit{et al.}, Phys. Rev. C 
\textbf{63}, 014003\ (2001).

\item D. E. Gonzalez Trotter, F. Salinas Meneses, W. Tornow, \textit{et al.}%
, Phys. Rev. C \textbf{73}, 034001\ (2006).

\item C. R. Howell, Q. Chen, T. S. Karman, \textit{et al.}, Phys. Lett. B 
\textbf{444}, 252\ (1998).

\item B. Gabioud, J.-C. Alder, C. Joseph, \textit{et al.}, Phys. Lett. B 
\textbf{103}, 9\ (1981).

\item I. \v{S}laus, Y. Akaishi, and H. Tanaka, Phys. Rev. Lett. \textbf{48},
993\ (1982).

\item H. Guratzsch, B. Kuhn, H. Kumpf, \textit{et al.}, Nucl. Phys. A 
\textbf{342, }239 (1980).

\item J. Soukup, J. M. Cameron, H. W. Fielding, \textit{et al.}, Nucl. Phys.
A \textbf{322, }109 (1979).
\end{enumerate}

\newpage

\vspace{1pt}\thispagestyle{empty} 
\noindent \textbf{Table.} Low-energy parameters of nucleon-nucleon
interaction in the $^{1}S_{0}$ state (the virtual-state energies in the
neutron-proton and proton-proton systems were calculated by formulas (3) and
(5); our results for the neutron-neutron system are given in the last
row)\bigskip

\begin{center}
\begin{tabular}{|c|c|c|c|}
\hline
$NN$ interaction & $\varepsilon _{NN}\,,\,$MeV & $a_{NN}\,,\,$fm & $%
r_{NN}\,,\,$fm \\ \hline
$np$ & $0.06640(27)$ & $-23.71(1)$ [4] & $2.70(9)$ [4] \\ 
$pp$ & $0.1211(19)$ & $-17.2(1)$ [4] & $2.82(10)$ \\ 
$nn$ & $0.1293(158)$ & $-16.59(117)$ & $2.83(11)$ \\ \hline
\end{tabular}
\\[0pt]
\end{center}

\end{document}